\documentclass[ 12pt]{iopart}
\usepackage{epsfig}
\def\be{\begin{equation}}
\def\ee{\end{equation}}
\def\ba{\begin{eqnarray}}
\def\ea{\end{eqnarray}}
\def\bL{\bar L}
\def\bC {\bar C}
\def\v{v_{dc}}

\begin{document}

\title{On Transmission Line Resonances in High T$_C$ dc SQUIDs}
\author{Urbasi Sinha $^1$ , Aninda Sinha $^2$ and Edward Tarte $^3$}
\address{$^1$ Institute for Quantum Computing, University of Waterloo, \\ 200 University avenue west, Waterloo,Ontario N2L 3G1,Canada} 
\address{$^2$ Perimeter Institute for Theoretical Physics, \\  31 Caroline street north, Waterloo, Ontario N2L 2Y5, Canada}
\address{$^3$ Department of Electrical and Electronic Engineering, University of Birmingham, Birmingham, UK}
\ead{$^1$ usinha@iqc.ca}
\ead{$^2$ asinha@perimeterinstitute.ca}
\ead{$^3$ e.tarte@bham.ac.uk}
\begin{abstract} In this paper, we study transmission line resonances in high $T_C$ dc SQUIDs. These resonances are exhibited in the characteristics of SQUIDs which are fabricated on substrates with high dielectric constant, such as strontium titanate. The power balance equation is analytically derived both for symmetric and asymmetric SQUIDs. Using this, we investigate SQUID current - voltage $I(V)$, voltage - flux $V(\Phi)$ and voltage modulation $\Delta V$ characteristics. 
\end{abstract}

\section{Introduction}
Superconducting quantum interference devices (SQUIDs) are the most sensitive magnetic flux sensors known today. Studying and improving SQUID properties has been a major research field in the last few decades \cite{clarke, gallop}. Most high $T_C$ SQUIDs are fabricated on strontium titanate (STO) substrates. It has been seen that large dielectric constant of STO strongly affects the performance of high $T_C$ dc SQUIDs by causing transmission line resonances to appear in their current-voltage characteristics \cite{Enpuku1,murayama,kuriki}. This was pointed out for instance, by Lee et al who experimentally showed that their hairpin shaped SQUID loop could be treated as a quarter wave microwave resonator in order to predict the voltage at which the resonance appeared in their measurements \cite{Lee}. Enpuku et al used a theoretical approach to predict the effect of the resonances based upon a transformation of the distributed resonator structure of the SQUID loop into a series of lumped element resonators \cite{Enpuku1}. This enabled them to calculate the current-voltage characteristics of the SQUID and later, the voltage-flux curves for other devices. However, it was necessary for them to solve the resulting differential equations numerically. In this paper, we develop an analytic approach to investigate the effects of transmission line resonances on dc SQUID resonance characteristics. We are able to obtain good agreement between experiment and theory for devices in the literature \cite{Lee}. We find that the resonance positions are mainly controlled by the product of the junction critical current $I_0$ and junction normal state resistance $R_S (I_0 \times R_S)$ as well as the dielectric constant of the substrate $\epsilon_R$.  We also investigate the effects of introducing asymmetry in junction parameters. We find that the resonance positions  are not very sensitive to the asymmetry. The actual magnitude of the current-voltage curve and $\Delta V $ curves are however sensitive to the degree of asymmetry introduced. We ignore the effects of thermal noise in performing our calculations. Noise analysis can be performed perturbatively as in \cite{noise}. In such calculations, the lowest order result is the noise free case where our analysis will be useful.

We begin with a brief outline of relevant previous work in the study of the effect of transmission line resonances on SQUID characteristics in section 2.  In section 3, we derive the power balance relation in terms of SQUID parameters, which involves deriving an expression for the circulating current through the SQUID inductance. Section 4 discusses current-voltage $I(V)$ curves for symmetric SQUIDs whereas in section 5, effects of introducing asymmetry on $I(V)$, voltage-flux $V(\Phi)$ curves and SQUID modulation-current $\Delta V(I)$ curves are discussed. The paper ends with a discussion of results obtained in section 6.

\section{ SQUID circuit equations}

\begin{figure}[b]
\begin{center}
\includegraphics[angle=-90,width=.9 \textwidth]{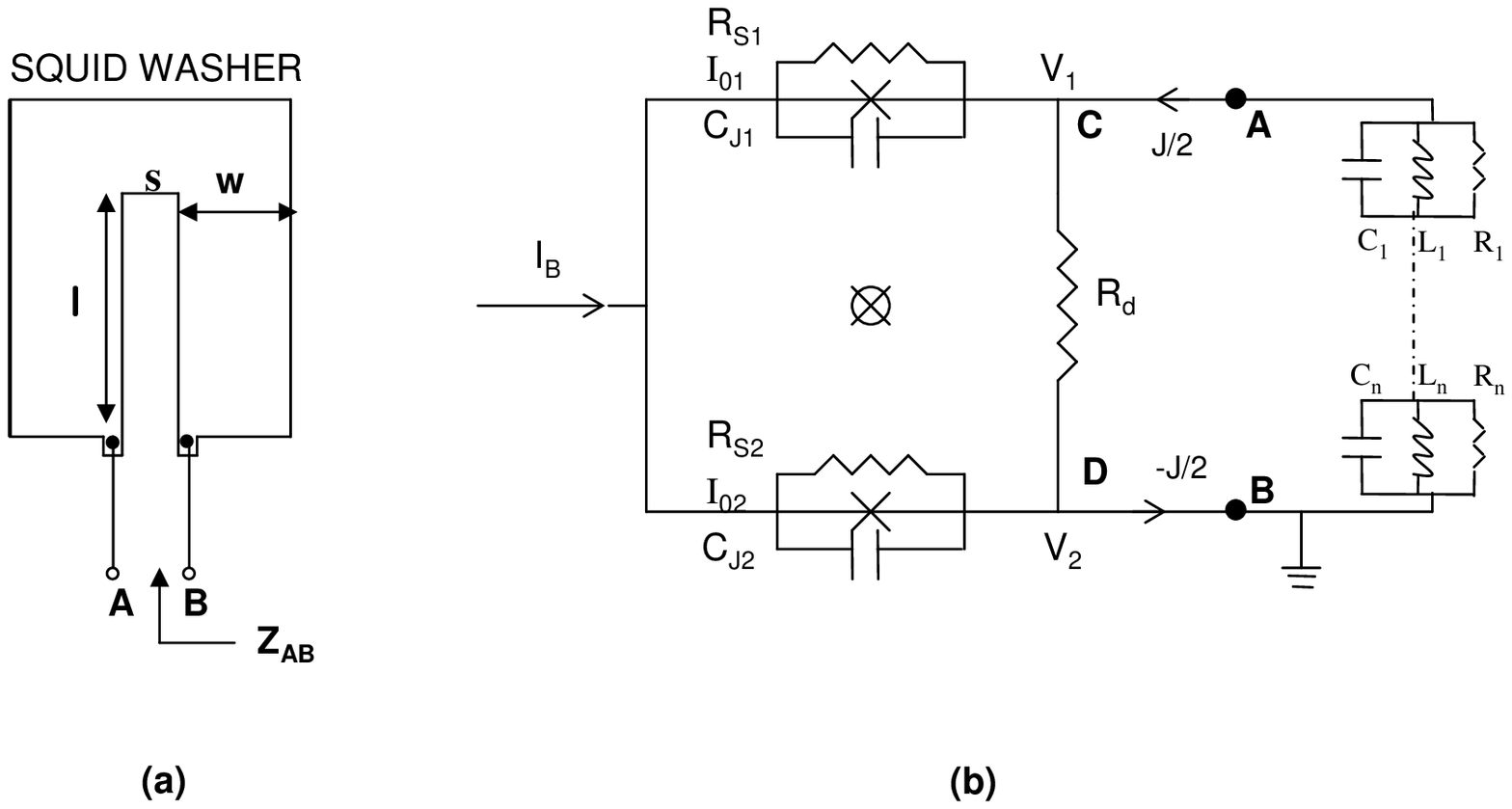}
\end{center} 
\caption{ (a) Geometry of the SQUID washer. Here {\it l}  is the length of the slit, {\it w} is the width of the electrode and {\it s} is the width of the slit.  (b) Equivalent circuit of the washer when its parasitic capacitance distributing along the slit of the washer is taken into account. The circuit consists of the SQUID coupled to a series of LCR circuits. $R_d$ is a damping resistance in parallel to the SQUID.}
\label{geometry}
\end{figure}

The geometry of the SQUID washer that we have used is shown in Fig.(\ref{geometry}a). This is a geometry commonly used to manufacture SQUIDs \cite{Lee, Ludwig, Bar} and is the geometry Enpuku et al used to numerically investigate the effects of large dielectric constant of strontium titanate (STO) on the  characteristics of high $T_C$ dc SQUIDs \cite{Enpuku1}. The slit of the SQUID washer makes up the SQUID inductance where $l$, $s$, $w$ and $d$ denote the slit length, slit width, electrode width and thickness of electrode respectively. For this geometry, the inductance per unit length of the slit $\bL$ and parasitic capacitance per unit length $\bC$ are given by \cite{Enpuku1, Yoshida}:
\be
\bL=\bL_{M}+\bL_{K}\,
\ee
where $\bL_M$ is the magnetic inductance per unit length and $\bL_K$ is the kinetic inductance per unit length of the SQUID slit given by\cite{Ramo}:
\be
\bL_{M} ={ \mu_0 K(k)\over K(k')}\,
\ee
\be
\bL_{K} = {2 \mu_0 \lambda^2 \over d w k'^2 K^2 (k')} [{w\over s} \ln [{4 w s \over d (w+s)}] + {w \over 2w + s}\ln [{4 w (2w+s)\over d (w+s)}]]
\ee
and
\be
\bC ={ \epsilon_{R} +1\over 2c^{2}\bL_{M}}
\ee
where $\mu_0$ is the permeability of free space, $K(k)$ is the complete elliptic integral of the first kind \cite{math} with a modulus $ k=\displaystyle {s\over s+2w}, k' = (1-k^2)^{1/2}, \lambda$ is the penetration depth of the film, $\epsilon_R$  is the dielectric constant of the STO substrate and $c$ is the velocity of light in vacuum.
The SQUID slit behaves as a transmission line with distributed inductance $L$ and distributed capacitance $C$. The impedance $Z_{AB}$ of the slit seen from terminals A and B is given by \cite{Enpuku1}:
\be\label{impedance}
Z_{AB}=iZ_{0}\tan (\Omega l \sqrt{\bL \bC}) + i\Omega L_{PR}\,
\ee
where $Z_0=\sqrt{\bL/\bC}$ is the characteristic impedance of the slit transmission line, $\Omega$ is the angular frequency of measurement and $L_{PR}$ is the junction parasitic inductance. The first term arises because the hairpin shaped slit can be treated as a shorted transmission line of length $l$. \\
Now, using the formula \cite{Enpuku1}
\be
\tan ({\pi x\over 2})=-{4x \over \pi} \displaystyle\sum_{n=1}^\infty {1\over x^2-(2n-1)^2}\,,
\ee
eqn.(\ref{impedance}) can be expanded as (in the lossless case with $L_{PR} = 0$):
\be\label{ZAB}
Z_{AB} = \displaystyle\sum_{n=1}^\infty {1\over i\Omega C_{n} + {1/ i\Omega L_{n}}}\,,
\ee
with
$C_n = {\bC l/ 2}$ and $L_n = {8\bL l/ \pi^2 (2n-1)^2}$. This transformation to an equivalent circuit has the advantage that it allows us to consider loss in the transmission line. In the lossy case, the rf-loss $R_n$ is added to eqn.(\ref{ZAB}) leading to:\\
\be\label{lossyZAB}
Z_{AB} = \displaystyle\sum_{n=1}^\infty {1\over i\Omega C_n + {1/ i \Omega L_n} + {1/ R_n}}\,,
\ee
where
$R_n = Q \sqrt{{L_n/ C_n}}$ and $Q$ is a quality factor{\footnote{It is possible to consider different quality factors for each $n$ but we will not do so here.}}. Here $l$ is the SQUID slit length. This expression allows us to express the impedance $Z_{AB}$ by the series of L-C-R resonant circuits as shown in fig.(\ref{geometry}b).

We will now show derivations of various quantities in the lossless limit and then quote the results obtained for the lossy case which can be derived analogously.  The circuit equations can be easily derived by the application of Kirchhoff's laws. In fig.(\ref{geometry}b), current entering point C should equal current leaving point C. We are assuming the most general case in which the SQUID is made up of junctions which are asymmetric \cite{Koelle, Ed}. Let the average junction critical current be $I_0$, the average junction normal state resistance be $R_S$ and the average junction capacitance be $C_J$. Let the asymmetry parameter in $I_0$ be $\kappa$, that in $R_S$ be $\rho$ and that in $C_J$ be $\chi$. So specifically, let us split \cite{Koelle} $2 I_0=(1+\kappa)I_0+(1-\kappa)I_0, {2/ R_S}={(1+\rho)/ R_S}+{(1-\rho)/R_S}$ and $2 C_J=(1+\chi)C_J+(1-\chi)C_J$. Then, this gives for the first junction:
\be\label{junction1}
(1+\kappa)I_0 \sin(\theta_1) + {V_1 (1+\rho)\over R_S} + (1+\chi)C_J {dV_1\over dt} - {I_B\over 2}  - {J\over 2} = {V_2 - V_1\over R_D}\,,
\ee
and writing an analogous equation for point D gives for junction 2:
\be\label{junction2}
(1-\kappa)I_0 \sin(\theta_2) + {V_2 (1-\rho)\over R_S} + (1-\chi)C_J {dV_2\over dt} - {I_B\over 2} + {J\over 2} = {V_1 - V_2\over R_D}\,,
\ee
Here $J$ is the circulating current through the SQUID inductance, $I_B$ is the SQUID bias current, $V_1$ and $V_2$ are voltages across junctions 1 and 2 , $\theta_2$ and $\theta_1$ are the phases of junctions 2 and 1 and $R_D$ is a damping resistance in parallel to the SQUID inductance. The circulating current in Fourier space is given by{\footnote{The expression for circulating current can be written explicitly in the frequency domain in terms of $V_1$ and $V_2$ which are obtained by solving eqns. (\ref{junction1}) and (\ref{junction2}). The equivalent expression for $j$ in the time domain (eqn.(B.22)) can only be explicitly obtained after solving the equations and taking into account the properties of the SQUID loop. This is presented in Appendix B.}}:
\be\label{j tilde}
\tilde J = 2{ \tilde V_2 - \tilde V_1 +{\Phi\over \bL l} \delta (\Omega)\over Z_{AB}}\,,
\ee
where $\delta$ is the Dirac delta function and $\Phi$ is the total flux applied to the SQUID loop.

We will follow \cite{Enpuku1} and use normalized circuit equations. We normalize currents by $I_0$, voltage by $I_0 R_S$ and time $t$ by ${\Phi_0/ 2\pi I_0 R_S}$. The ac Josephson relation gives  ${\it v} = {d\theta / d \tau} = {V/I_0 R_S}$ , where $v$ is the normalised voltage and ${\it \tau}$ is the normalised time. Here we have neglected random noise currents. The normalized circuit equations are:
\be\label{normalised1}
(1+\chi)\beta_C \ddot\theta_{1} = {1\over 2} (i_B + j) - (1+\rho)\dot\theta_{1} - (1+\kappa)\sin (\theta_1) - \gamma (\dot\theta_{1} - \dot\theta_{2})\,,
\ee
\be\label{normalised2}
(1-\chi)\beta_C \ddot\theta_{2} = {1\over 2} (i_B - j) -(1-\rho) \dot\theta_{2} - (1-\kappa)\sin (\theta_2)  +\gamma (\dot\theta_{1} - \dot\theta_{2})\,,
\ee
where  $\beta_C = {2 \pi I_0 C_J R_{S}^2/ \Phi_0}$ is the SQUID McCumber parameter with $\Phi_0$ being the flux quantum and $ \gamma = {R_S/ R_D}$. In principle, these equations can be solved numerically using the dielectric constant $\epsilon_R$  as a fit parameter having a finite number of resonant circuits. Enpuku et al present numerical solutions of eqns.(\ref{normalised1}) and (\ref{normalised2}) in their paper. As shown in the appendix A, they however consider a finite number `$n$' of resonant circuits and the expression they use for the circulating current $J$ becomes indeterminate in the continuum limit.  
The analytic handle will enable closed form expressions for $I_{B}(V)$, which can subsequently be used to simulate graphs quickly and conveniently. Normalizing eqn.(\ref{j tilde}) and the expressions for impedance in the lossless limit, eqn.(\ref{ZAB}) and lossy limit, eqn.(\ref{lossyZAB}),  the expression for the normalized circulating current $j$ in Fourier space is given by:
\be\label{normalised jtilde}
\tilde j = ( \tilde \theta_2 - \tilde \theta_1 + { 8 \pi \over  \beta} \delta (\omega) \phi ) A (\omega)
\ee
where $\omega = {\Omega \Phi_{0}/ 2 \pi I_0 R_S}$ and is dimensionless. Here $\phi$ is the externally applied flux normalized to $\Phi_0$ and $ \beta = { 2 \bL l I_0/ \Phi_0}$ is the SQUID inductance parameter. In the lossless case with $L_{PR} = 0$
\be \label{Anoloss}
A (\omega) ={2 \over \bL l } {\Phi_0 \over 2 \pi I_0} \frac{ ({\omega l 2 \pi I_0 R_S \sqrt{ \bL \bC}/\Phi_0})}{ \tan ({\omega l 2 \pi I_0 R_S \sqrt {\bL \bC}/\Phi_0})}
\ee
and in the lossy case, performing the sum in eqn.(\ref{lossyZAB}):
\be\label{Aloss}
A(\omega)=2i {\nu\over\mu} [\Psi ({\bf \zeta}-i {\bf \xi})+\Psi ({\bf \zeta}+i {\bf \xi})]\,,
\ee
where
$ {\bf \zeta}={1/ 2}+{i\omega l\sqrt{\bL \bC}/ 2\pi Q}$, ${\bf \xi}={i l\omega \sqrt{\bL\bC (1-4 Q^2)}/ 2\pi Q}$, $\nu=-{Q \Phi_0/ 2\pi I_0 \sqrt{\bL l}}$, $\mu=\bL\pi\omega\sqrt{l\bC (1-4Q^2)}$ and $\Psi(x)= \partial_x \log \Gamma(x)$ is the di-gamma function with $\Gamma(x)$ being the standard gamma function \cite{math}. If $L_{PR}\neq 0$ then $A(\omega)$ should be replaced by ${A(\omega)/ (1+{\pi I_0 A(\omega) L_{PR} / \Phi_0})}$ in the calculations.

\section { Power balance equation}
The calculation details for the SQUID power balance equation have been included in Appendix B. After some straightforward but tedious algebra, the power balance relation so derived is as follows:
\begin{eqnarray} \label{exact}
i_b&=&2\v + {I^2 \v \over 2 |d s-r \bar r|^2} \bigg\{ (1+\kappa^2) [d \bar d+r \bar r] +2 \kappa [d \bar r +\bar d r]\nonumber \\  &+&(1-\kappa^2)\bigg( [d \bar d-r \bar r]\cos (2\pi\phi) 
   + i [d \bar r-\bar d r ]\sin(2\pi\phi)\bigg)\nonumber \bigg\} \\
   &+&  {I^2 \v \over 2 \beta^2 |d s - r \bar r|^2} (1+2\gamma- i{A-\bar A\over 2 \v})   \bigg\{ (1+\kappa^2) [\beta^2 s \bar s + r \bar r] - 2\kappa\beta [\bar s r + s \bar r]\nonumber \\ &-&(1-\kappa^2)\bigg ([\beta^2 s \bar s - r \bar r]\cos(2\pi\phi) - i\beta [s \bar r - \bar s r ]\sin (2\pi\phi)\bigg )\bigg\}\,.
\end{eqnarray}
Here $d = d(v_{dc}), s = s(v_{dc})$, $ r = r(v_{dc})$, $A=A(\v)$ and $\bar x$ denotes the complex conjugate of $x$. The phase difference between the two junctions have been set to $2\pi\phi$.
Here,
\be
I = \sqrt { 2 v_{dc} ( 1 + v_{dc}^2 )^{1/2} - v_{dc} }\,
\ee
\be\label{sv}
s(\omega) = ( \omega^2 \beta_C + i \omega )\,, \quad d(\omega)= A(\omega)-(\omega^2 \beta_C + i \omega + 2 i \omega \gamma)\,,
\ee
\be
r(\omega)=i\rho\omega+\chi\beta_C\omega^2\,,
\ee

Note that the power balance relation has the correct symmetries. For instance interchanging the two junctions leads to $\kappa\rightarrow -\kappa$, $\rho\rightarrow -\rho$, $\chi\rightarrow -\chi$ and $\psi_1\leftrightarrow \psi_2$. The expression is symmetric under this transformation. Another interesting observation is the fact that the $\rho$ and $\chi$ dependence are both packaged in the function $r(\omega)$.

\section {Symmetric SQUIDs}

For symmetric SQUIDs, i.e. $ \kappa = \rho =\chi= 0$ and in the lossless case where $A=\bar A$, we have
\begin{eqnarray} \label{symmetric}
i_b&=&2\v+ {I^2 \v \over 2 |s|^2}\bigg\{1+ \cos 2\pi \phi \bigg\}
   +  {I^2 \v \over 2  |d|^2} (1+2\gamma)  \bigg\{ 1  -\cos 2\pi \phi \bigg\}\,,\nonumber\\
\end{eqnarray}
where $s,d$ are given by eqn.(\ref{sv}) and $A$ is given by eqn.(\ref{Anoloss}). In reference \cite{Enpuku2}, Enpuku et al derive the SQUID power balance equation in the absence of transmission line resonances. By allowing the slit length to tend to zero, hence treating the SQUID inductance as a lumped inductance instead of a distributed parameter, we retrieve Enpuku's expression for the power balance relation:\\
\be\label{Enpuku}
i_B v_{dc} = 2 v_{dc}^2 + {I^2 (1+ \cos(2 \pi\phi)) \over 2 (1+ \beta_{C}^{2} v_{dc}^2)} + {I^2 ( 1- \cos(2\pi\phi)) (1+ 2\gamma) \beta^2 v_{dc}^2 \over 2 ( [ {2\over \pi} - \beta \beta_C v_{dc}^2 ]^2 + \beta^2 v_{dc}^2 ( 1+ 2\gamma)^2)}
\ee
Here we have neglected the capacitance in parallel to the shunt resistance, which they had included in their circuit as our analysis is based on a circuit, which does not involve such a capacitance term. 
\begin{figure}[ht]
\begin{tabular}{ll}
\hskip 2cm \includegraphics[width=.45 \textwidth]{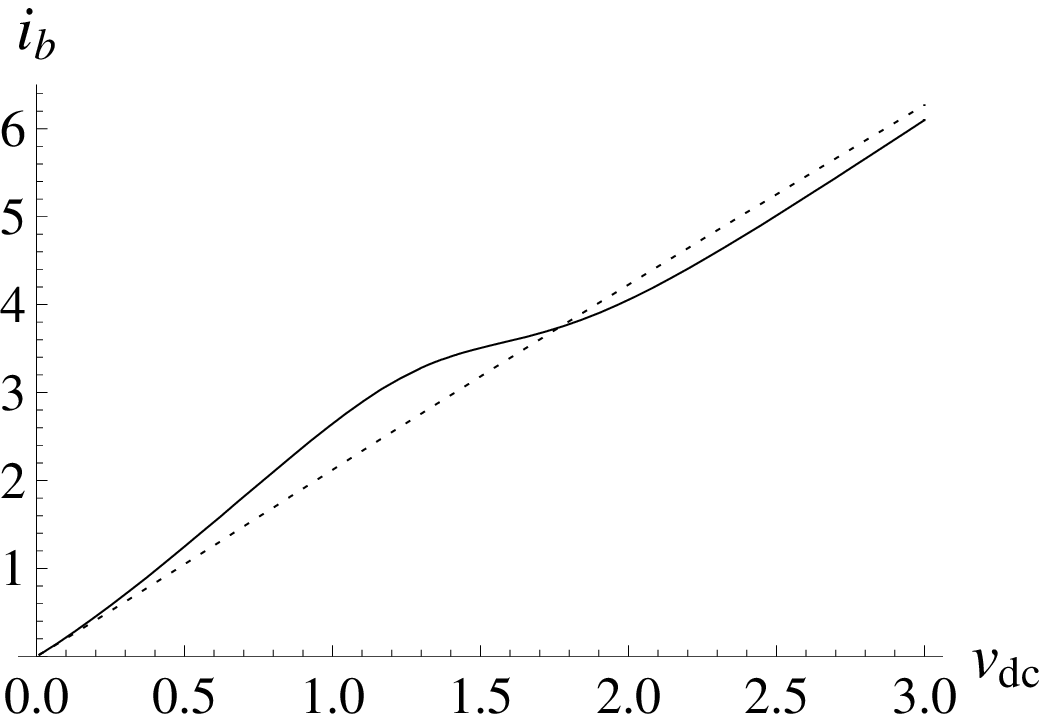}
&\includegraphics[width=0.45 \textwidth]{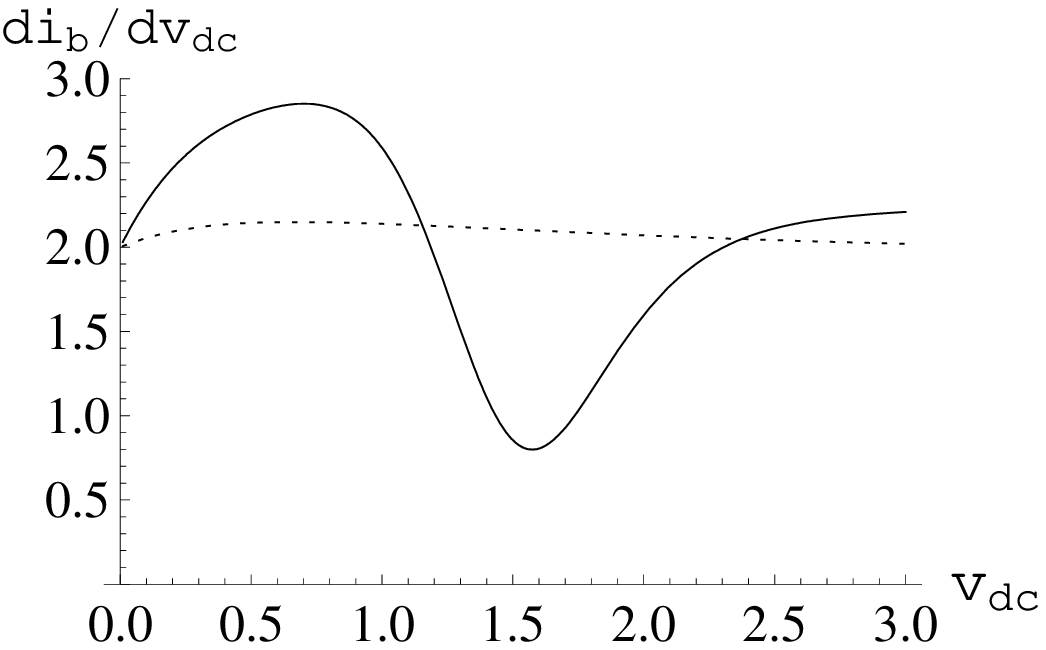}\\
\hskip 5cm (a)& \hskip 3cm (b)
\end{tabular} 
\caption{(a) Comparison between the $i_b (v_{dc})$ curve from eqn.(\ref{exact}) which takes transmission line resonances into account (solid black line) and the $i_b (v_{dc})$ curve from the equation derived by Enpuku et al, eqn.(\ref{Enpuku}) which considers the SQUID as a lumped element (dotted line) for $\phi =1/2$ using Lee et al's parameters. The step in the solid black line at around $v_{dc} = 1.2$ represents a resonance position. Such features are clearly absent in the dotted line. (b) Comparison between the corresponding $di_b / dv_{dc} (v_{dc})$ curves. The solid black line is derived from eqn.(\ref{exact}) and the dotted line from eqn.(\ref{Enpuku}).}
\label{Enpukucomp}
\end{figure}
Lee et al \cite{Lee} had seen a current step in their $I(V)$ curve which they attributed to the transmission line resonances due to the high dielectric constant of the substrate. We have used their SQUID parameter values ($l=55\mu m$, SQUID inductance, $L_{SQ}(\bar L \times l) = 55pH, I_0 = 4.75 \mu A, R_S = 13.8\Omega$, SQUID parasitic inductance $L_{PR}= 13pH$ and $\beta_{C} = 0$) to simulate $i_b(v_{dc})$ curves from eqn.\ref{exact} in the lossless case. In accordance with Lee et al, $\epsilon_R = 1930$. They had seen the resonant current step at $80\mu V$. 
Fig.(\ref{Enpukucomp}a) shows the comparison between the curves we get from eqn.(\ref{exact}) and eqn.(\ref{Enpuku}) respectively. Fig.(\ref{Enpukucomp}b) shows the comparison between the differential conductance $di_{b}/dv_{dc} (v_{dc})$ curves obtained from differentiating eqn.(\ref{exact}) and eqn.(\ref{Enpuku}) respectively. Enpuku et al's formula does not lead to any resonances which clearly arise from the consideration of the SQUID loop behaving as a transmission line. The minima in the $i_b(v_{dc})$ curve occur for $A\rightarrow \infty$ whereas the maxima occur for $A\rightarrow 0$.

Eqn.(\ref{exact}) is the SQUID power balance equation when transmission line resonances due to the distribution of the parasitic capacitance due to high dielectric constant of the STO substrate along the slit of the SQUID washer are taken into account. The resonances only affect the terms of this equation which contains $A(\omega)$. This is because these terms are associated with $D=\theta_1-\theta_2$ and hence with the circulating current. Thus, where $\cos 2\pi\phi =1$, ie for $\Phi_{ext} = n\Phi_0$ the resonances have no effect on the current-voltage characteristics, whereas for $\cos 2\pi\phi = -1, \Phi_{ext} = {1 \over 2} (2m+1)\Phi_0$,  the resonances have maximum effect. \\
The maxima occur at:\\
\be
\v= {m \Phi_0 \over 2l I_0 R_S} {1\over \sqrt{\bL\bC}}
\ee
The minima occur at:\\
\be\label{reso}
\v = {(2n+1) \Phi_0 \over 4l I_0 R_S} {1 \over \sqrt {\bL\bC}}
\ee

If we approximate $ \sqrt {\bL\bC}$ by $ \sqrt {{( \epsilon_R + 1)\over 2c^2}} $, where $c$ is the speed of light in free space, then eqn.(\ref{reso}) is the one used by Lee et al to predict the voltage at which the resonance occurred for their device. This approximation is only true if the kinetic inductance of the SQUID loop is negligible. \\

\section{Introducing asymmetry}
\subsection{Current-voltage characteristics}
\begin{figure}[ht]
\begin{tabular}{ll}
\hskip 2cm \includegraphics[width=0.45 \textwidth]{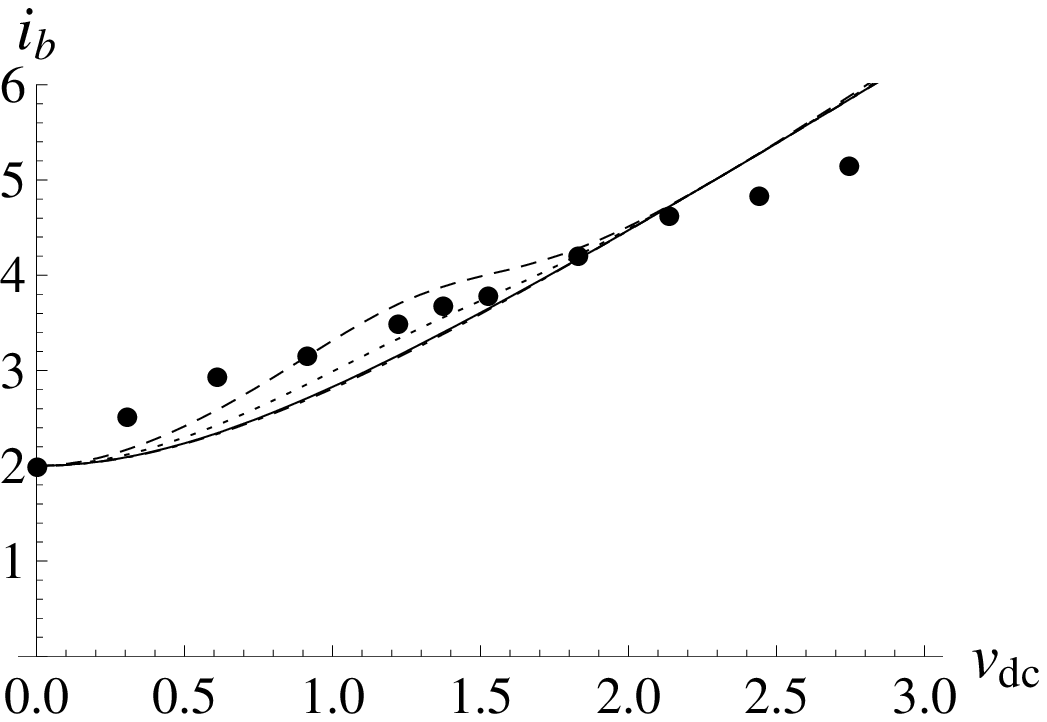}
&\includegraphics[width=0.45 \textwidth]{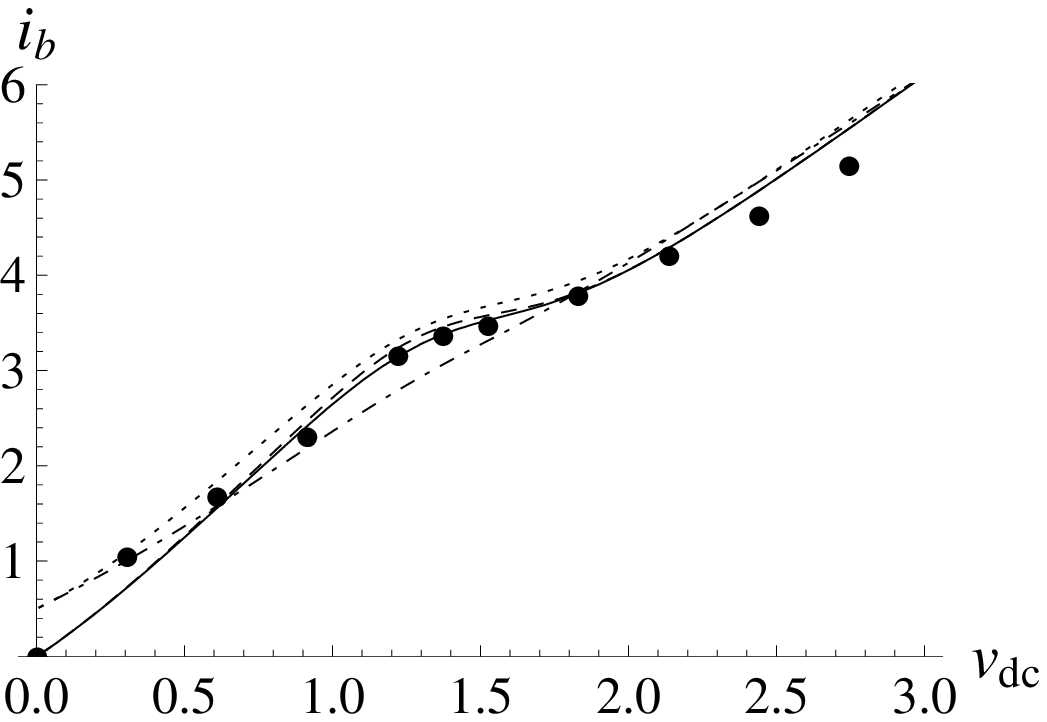}\\
\hskip 5cm (a)& \hskip 3cm (b)
\end{tabular} 
 {\caption{(a) Comparison between Lee et al's experimental data read out from \cite{Lee} (bold black dots) and theoretical simulations from eqn.(\ref{exact}) for the $\phi = 0 $ case. Continuous line is the simulated curve for no asymmetry considered, dotted line is for the case when current asymmetry is considered and $\kappa = 0.5$, the dashed line is for the case when normal state resistance asymmetry is considered and $\rho = 0.2$ and the dot-dashed line is for the case when both current and resistance asymmetry are considered and $\kappa = 0.5$ and $\rho = 0.2$. (b) The $\phi = 1/2$ case. } \label{Leezerohalf}}
\end{figure}

In this section we wish to consider the effects of introducing asymmetry to the SQUID.\\
We have introduced current and resistance asymmetry systematically and seen the effect on the $I(V)$ curve both for the $\phi = 0$ and $\phi = 1/2$ cases. Fig.(\ref{Leezerohalf}a) shows the comparison between Lee's experimental $I(V)$ and our theoretical simulations for the $\phi=0$ case and fig.(\ref{Leezerohalf}b) shows the same for the $\phi =1/2 $ case.  
The comparison between experiment and theory in the $\phi = 0$ case is not very good at large voltages. In the $\phi = 1/2$ case, introducing current asymmetry shifts the origin to a point higher than zero as expected. Furthermore, in the $\phi =1/2$ case, the curve with no asymmetry seems to be a good fit to the experimental data. When both current and resistance asymmetries are considered, it leads to a general flattening of the curves, leading to very broad resonances. 
 We have chosen $\kappa = 0.5$ and $\rho = 0.2$ as typical asymmetry parameters. We find that increasing $\rho$ to a value higher than 0.2 leads to extremely flattened $i_b (v_{dc})$ curves and also multiple roots in the $v_{dc} (\phi)$ curves (next section)  which indicates that 0.2 is probably the highest resistance asymmetry we can consider in this case. We have also chosen to demonstrate trends for $\kappa = 0.5$ as we feel that for practical SQUIDs this represents one of the highest asymmetries.

\subsection {Voltage-Flux characteristics}
\begin{figure}[ht]
\begin{tabular}{ll}
\hskip 2cm \includegraphics[width=.45 \textwidth]{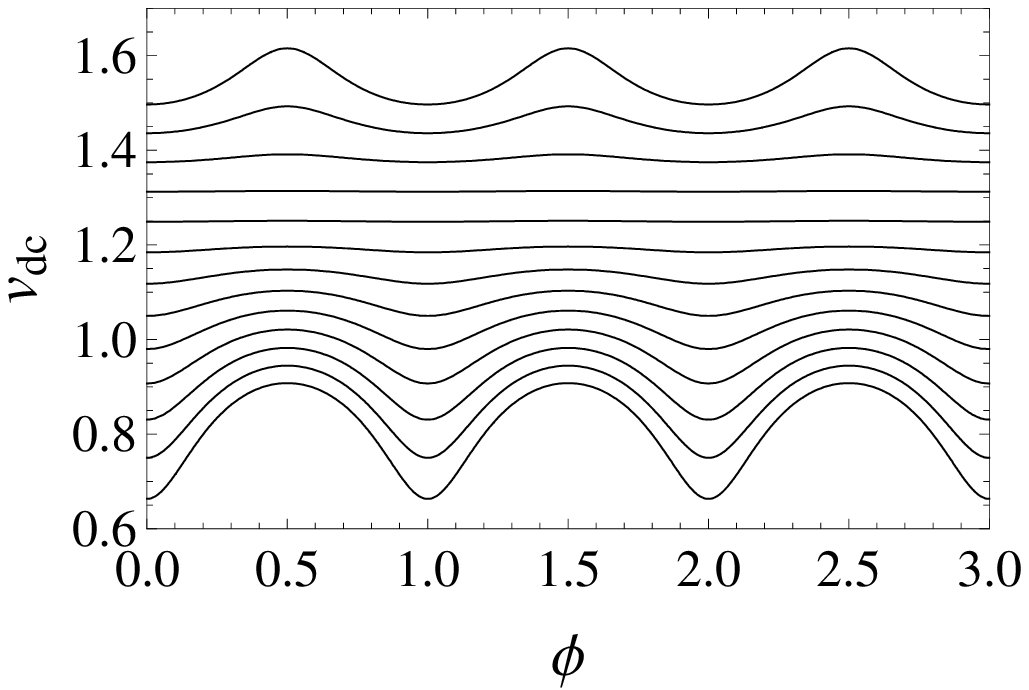}
&\includegraphics[width=.45 \textwidth]{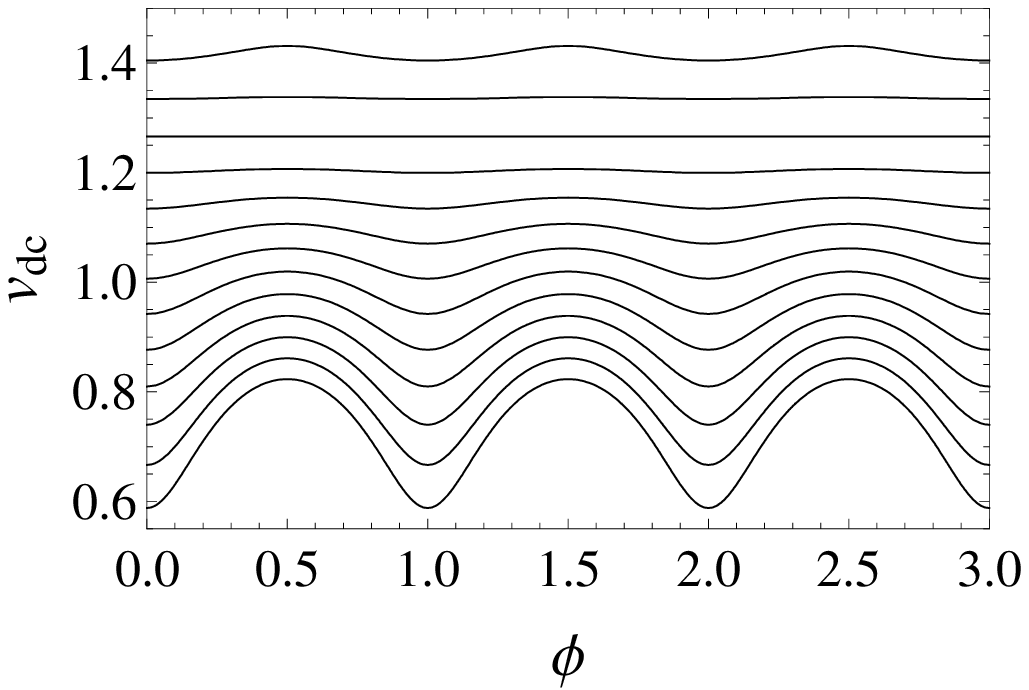}\\
\hskip 5cm (a)& \hskip 3cm (b)
\end{tabular} 
\caption{(a) $v_{dc} (\phi)$ curve in the absence of any asymmetry in SQUID parameters. (b) $v_{dc} (\phi)$ curve in the presence of current asymmetry $\kappa = 0.5$. The minima exhibit resonances and go to zero. In both the figures, bottom curve is for $i_b = 2.4$ and top curve is for $i_b = 3.6$.}
\label{vphi1}
\end{figure}

Eqn.(\ref{exact}) can also be used to simulate the voltage-flux  $v_{dc}-\phi$ characteristics of a SQUID. We have again used Lee et al's  SQUID parameter values to simulate $v_{dc}-\phi$ curves for various cases viz. in the absence of any asymmetry in SQUID parameters, in the presence of just current asymmetry, in the presence of just resistance asymmetry and also in the presence of both current and resistance asymmetries. \\
Fig.(\ref{vphi1}a) shows the $v_{dc}-\phi$ curve in the absence of asymmetries. Fig.(\ref{vphi1}b) shows the $v_{dc}-\phi$ curve in the case when a current asymmetry $\kappa = 0.5$ is considered.

\begin{figure}[hb]
\begin{tabular}{ll}
\hskip 2cm \includegraphics[width=.45 \textwidth]{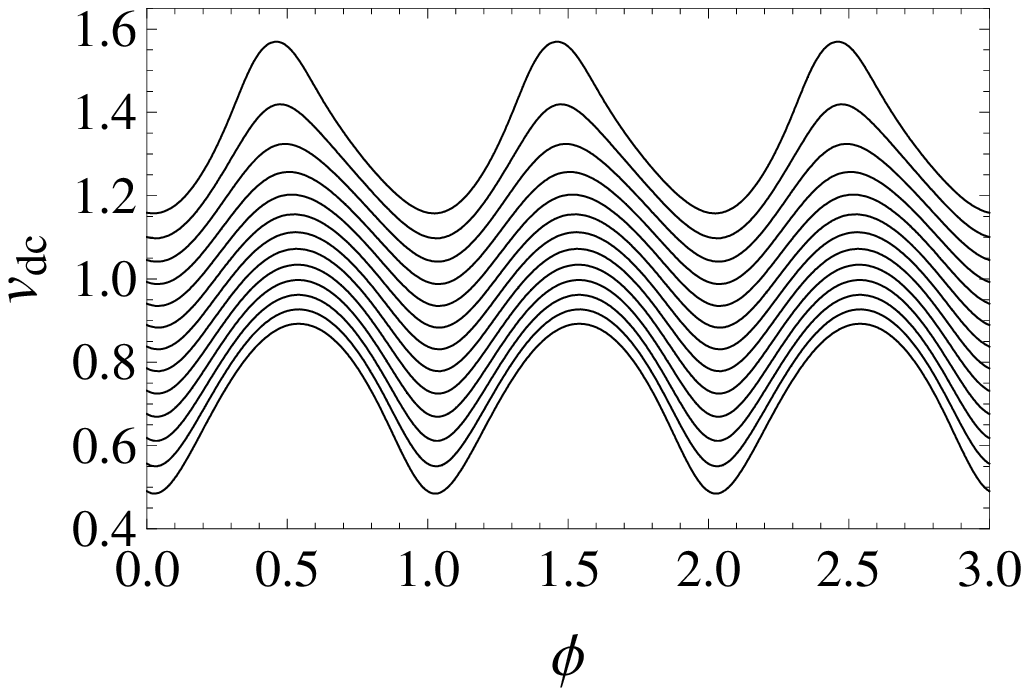}
&\includegraphics[width=.45 \textwidth]{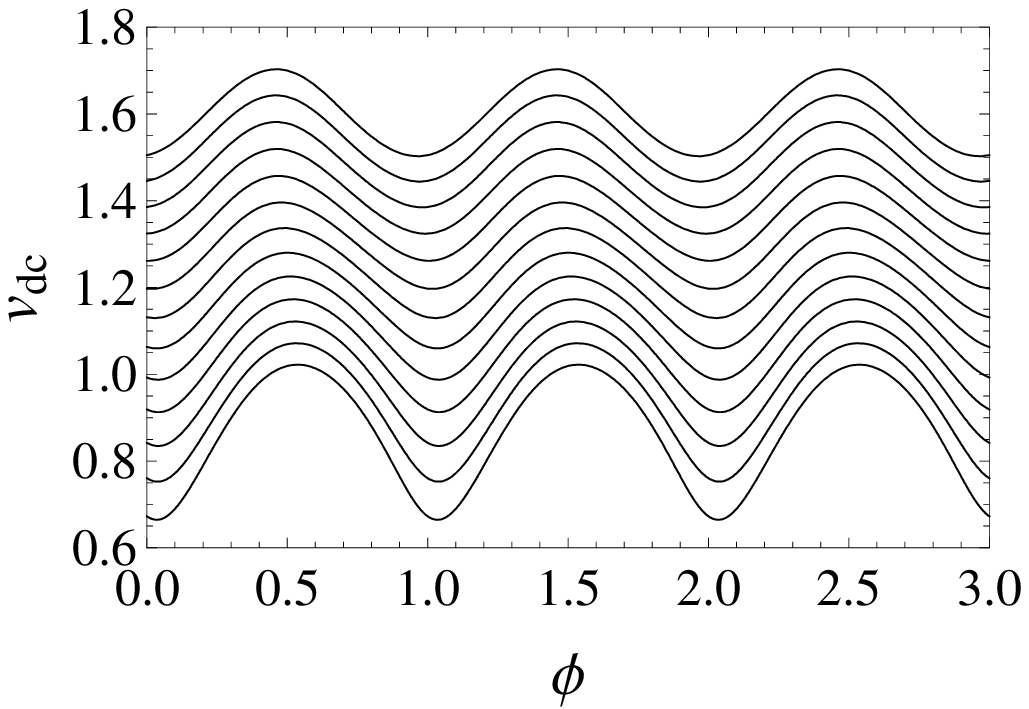}\\
\hskip 5cm (a)& \hskip 3cm (b)
\end{tabular} 
\caption{(a) $v_{dc} (\phi)$ curve in the presence of resistance asymmetry $\rho = 0.2$. (b) $v_{dc} (\phi)$ curve in the presence of current asymmetry $\kappa = 0.5$ and resistance asymmetry $\rho = 0.2$. The minima which indicate resonance positions do not go to zero anymore. In both the figures, bottom curve is for $i_b = 2.4$ and top curve is for $i_b = 3.6$.}
\label{vphi2}
\end{figure}

Fig.(\ref{vphi2}a) shows the $v_{dc}-\phi$ curve in the case when a normal state resistance  asymmetry $\rho = 0.2$ is considered. Fig.(\ref{vphi2}b) shows the $v_{dc}-\phi$ curve in the case when a current asymmetry $\kappa = 0.5$ and resistance asymmetry $\rho =0.2$ are considered. In each of the four cases, normalized bias current $i_b$ ranges from 2.4 to 3.6.

From fig.(\ref{vphi1}), we find that in the absence of any asymmetry we have perfectly symmetric $v_{dc} (\phi)$ curves. The difference between the maximum and minimum $v_{dc}$ decreases i.e. $\Delta v_{dc}$ decreases for increasing values of bias current, $\Delta v_{dc}$ goes to zero for $i_b = 3.3$. Then it again increases for higher bias current values. The complete flattening of the $v_{dc} (\phi)$ curve or $ \Delta v_{dc}$ going to zero indicates a resonance position. With the introduction of current asymmetry (fig.(\ref{vphi1})), the voltage  values are less than the corresponding ones in the absence of asymmetry for a particular bias current; for instance, for $i_b = 3.6$, maximum $v_{dc}$ in the absence of asymmetry is around 1.62 whereas for the same bias current the maximum voltage decreases to around 1.45 in the presence of current asymmetry. Both fig.(\ref{vphi2}a), where only resistance asymmetry is considered and fig.(\ref{vphi2}b), where both resistance and current asymmetries are considered  exhibit skewed $v_{dc} (\phi)$ curves. Beyer et al \cite{Beyer} found that $v_{dc} (\phi)$ curves got skewed on introduction of asymmetry in SQUID parameters. We find that  complete flattening of $v_{dc} (\phi)$ curves is absent when resistance asymmetry is introduced i.e. the minima in the curves do not go to zero. From Lee's experimental data ( for example see fig.(\ref{Leezerohalf}a,b)) we find, that at $i_b=3.3$, the voltage values for $\phi=0$ and $\phi=1/2$ are almost the same. This indicates a resonance position. Thus we expect flattening of the voltage-flux curves at this position. This indicates that even though some current asymmetry may be present in the SQUID (see fig.(\ref{vphi1}b)), resistance asymmetry is probably absent.
\subsection { Voltage modulation $\Delta V$}
\begin{figure}
\begin{center}
\includegraphics[width=.75 \textwidth]{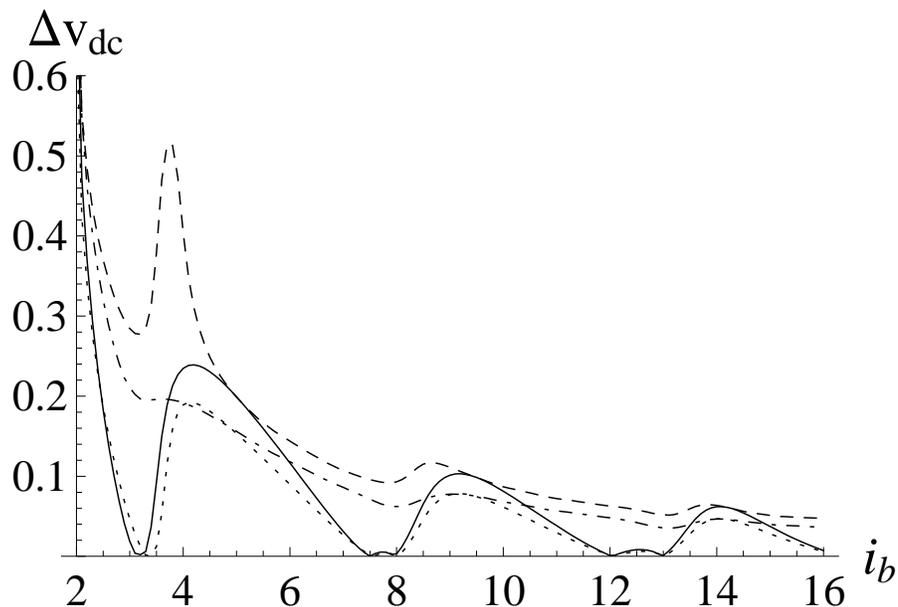}
\end{center} 
\caption{ Comparison between voltage modulation $\Delta v_{dc} (i_{b})$ where both $\Delta v_{dc}$ and $i_{b}$ are in normalized units in cases where no asymmetry is considered and when various asymmetries are introduced. The continuous line is the simulated $\Delta v_{dc} $ for no asymmetry considered, the dotted line is for the case when current asymmetry is considered and $\kappa = 0.5$, the dashed line is for the case when normal state resistance asymmetry is considered and $\rho = 0.2$ and the dot-dashed line is for the case when both current and resistance asymmetry are considered and $\kappa = 0.5$ and $\rho = 0.2$. }
\label{deltav}
\end{figure}

Fig.(\ref{deltav}) shows the comparison between voltage modulation $\Delta v_{dc} (i_{b})$ where both $\Delta v_{dc}$ and $i_{b}$ are in normalized units in cases where no asymmetry is considered and when various asymmetries are introduced. 

We find that the resonance position is almost unaffected when just current asymmetry is introduced. However, the magnitude of $\Delta v_{dc}$ decreases by around 20\%. On introduction of just resistance asymmetry on the other hand, the magnitude of $\Delta v_{dc}$ increases, by almost 100\% for the first peak and by around 20\% for the next couple of peaks. The resonance position also seems to shift towards the left (towards smaller current values). On introduction of both current and resistance asymmetries, the resonances get rather flattened and there is a general decrease in magnitude of $\Delta v_{dc}$ in keeping with the observation for just current asymmetry and the resonance position, especially the first peak shifts a little leftward in keeping with the observation for just resistance asymmetry. The other peaks do not show any significant shift.\\

\section {Summary of results}
In this paper, we analytically investigated the effects of transmission line resonances on dc SQUID characteristics. We obtain a closed form expression for the $i_b(\v)$ characteristics of the SQUID. We use SQUID parameters used by Lee et al \cite{Lee} and compare our theoretical simulations with their experimental data. We find that the resonance positions are mainly controlled by $I_0 R_S$ as well as the dielectric constant of the substrate $\epsilon_R$.  We also investigated the effects of introducing asymmetry in junction parameters. We find that the resonance positions  are not very sensitive to the asymmetry. The actual magnitude of the current-voltage curve and $\Delta V $ curves are however sensitive to the degree of asymmetry introduced. Our analytical solution may provide us with a method of theoretically determining the value of the dielectric constant of the substrate if we use that as a fit parameter between experimental and theoretical graphs. In an  earlier work, we have determined $\epsilon_R$ of STO thin films using a Josephson junction based technique called Josephson Broadband Spectroscopy \cite{JOBS1, JOBS2}.

\ack{We wish to thank John Gallop for his enthusiasm about the calculations performed and useful suggestions and Frank Wilhelm for his useful comments and suggestions.}

\appendix 
\section {}
Enpuku et al's  expression for the circulating current $J$(eqn.(12) of \cite{Enpuku1}) is given by:
\be
J = j_0 (\theta_2 - \theta_1 + \displaystyle\sum_{n=1}^\infty {m_n \theta_{r,n}} ) + {4 \over \beta} \phi_{ex}\,
\ee
where $m_n = M_n / L_n$, $M_n$ is the mutual inductance between SQUID and $n^{th}$ resonant circuit, $L_n$ is the inductance of the $n^{th}$ resonant circuit, $\theta_{r,n}$ is the phase corresponding to the voltage $V_{r,n}$ across $L_n$, $ \phi_{ex}$ is the external applied flux and $ j_0 = {2 \over \pi \beta} \bigg({1\over 1 -  \displaystyle\sum_{n=1}^\infty {\alpha_{n}^{2}}}\bigg)$ where $\beta$ is the SQUID inductance parameter and $\alpha_n$ is the coupling constant between the $n^{th}$ resonant circuit and the lumped SQUID loop given by $ \alpha_n = {2 \sqrt{2} \over \pi (2n-1)}$. A drawback in the expression of $J$ is immediately apparent. Substituting the expression for $\alpha_n$, we get for $J$:\\
\be
J = {1 \over ( 1 - \displaystyle\sum_{n=1}^\infty {\alpha_{n}^{2}})} {1 \over \pi \beta} ( \theta_2 - \theta_1 + \displaystyle\sum_{n=1}^\infty {m_n \theta_{r,n}}) + {4 \phi_{ex}\over \beta}
\ee
There is a problem with this expression in the continuum limit when $n \rightarrow\infty$. 
In this limit, 
\be
\theta_2 - \theta_1 = - \displaystyle\sum_{n=1}^\infty {m_n \theta_{r,n}} \,
\ee
$\Rightarrow J = \frac{0}{0} + {4 \phi_{ex}\over \beta}$,  , which is undefined. 
 
 \section {}
 In the following derivation of the SQUID power balance relation, random noise currents have been neglected.
 Multiplying eqn.(\ref{normalised1}) by $\dot\theta_{1}$ and eqn.(\ref{normalised2}) by $\dot\theta_{2}$, adding the two and noting that the voltage across the SQUID averaged over a time $T$ is given by:
\be
{1\over T} \displaystyle\int^T_0 \dot\theta_{1} dt = v_{dc} = {1 \over T}  \displaystyle\int^T_0 \dot\theta_{2} dt \,
\ee
we obtain:
\be \label {ibvdc1}
v_{dc} i_B = \displaystyle\lim_{T\to\infty} {1 \over T} \displaystyle\int^T_0 dt ( \dot\theta_{1}^2 + \dot\theta_{2}^2 + \gamma ( \dot\theta_{1} - \dot\theta_{2} )^2 - {1\over 2} j ( \dot\theta_{1} - \dot\theta_{2}))\,
\ee
Eqn.(\ref {ibvdc1}) can be rewritten as:
\be \label {ibvdc2}
i_B v_{dc} = \displaystyle\lim_{T\to\infty}{ 1 \over T} \displaystyle\int^T_0 {1\over 2 } ( \dot S^2 + ( 1 + 2 \gamma ) \dot D^2 - j (t) \dot D) dt\,
\ee
where $ S = \theta_1 + \theta_2 $ and $ D = \theta_1 - \theta_2 $.
The left hand side of eqn.(\ref{ibvdc1}) represents the input power from the bias current $i_B$. The first two terms on the right hand side of eqn.(\ref{ibvdc1}) represent the power dissipated in the two SQUID shunt resistances, the third term represents the power dissipated in the damping resistance while the fourth term represents the power dissipated in the resonant circuit. The time averaged voltage as a function of bias current and applied flux $V(I_B, \Phi )$  for the SQUID can be obtained if the voltage waveforms for the two junctions $v_1$ and $v_2$ are known. When a dc voltage $\v$ appears across the SQUID, the normalized Josephson currents, $\sin(\theta_1)$ and $\sin(\theta_2)$  oscillate with time since the relation  
$$ 
 \sin (\theta_i) = \sin ( \displaystyle\int v_i dt) \,
 $$ 
 holds, where $i=1,2$. Moreover, by averaging   over the period of a Josephson oscillation, it is simple to show that the fundamental frequency of the Josephson current may be given by ${v_{dc}/ 2 \pi}$. Therefore, we can express the Josephson current in the presence of $v_{dc}$ as:
\be
\sin (\theta_i) = \displaystyle\sum_{n=1}^\infty i_{i,n} \cos ( n v_{dc} t + \psi_{i,n})\qquad i=1,2\,
\ee
where $i_{i.n}$ and $\psi_{i,n}$ represent the amplitude and phase of the $n^{th}$ Fourier mode respectively. The above relation means that the Josephson current can be regarded as an ac current generator in the presence of $v_{dc}$. The amplitude of the current generator is given by \cite{Enpuku2}:
\be
I = i_1 = i_2 = \sqrt { 2 v_{dc} ( 1 + v_{dc}^2 )^{1/2} - v_{dc} }\,
\ee
The phases of the two current generators are different and the phase difference is given by the normalized applied flux as $\psi_1 - \psi_2 = 2 \pi \phi$ . In the present calculation, only the first harmonic of the Fourier expansion has been considered i.e.  $ \sin (\theta_i) = i_i \cos ( v_{dc} t + \psi_i) $ has been used.
We have found that this approximation is in good agreement with numerical simulations as well over the Josephson frequency range. In this paper, the following definition has been used for the Fourier transform:
\be
 \tilde X (\omega) = \int^{\infty}_{-\infty} X(t) e^{i \omega t} dt\,
 \ee
 and 
 \be
 X (t) = {1\over 2\pi} \int^{\infty}_{-\infty} \tilde X(\omega) e^{-i \omega t} d\omega\,
 \ee
 Taking Fourier transform of eqn.(\ref{normalised1}) and eqn.(\ref{normalised2}) gives:
 \ba
 &-& (1+\chi)\omega^2 \beta_{C} \tilde \theta_{1} \nonumber \\ &=& {1\over 2} i_{B} \delta (\omega) + {1\over 2} \tilde j + i \omega ( 1 + \rho ) \tilde \theta_{1} - ( 1 + \kappa ) \widetilde{ \sin(\theta_{1})} + i \omega \gamma ( \tilde \theta_{1} - \tilde \theta_{2} )
 \ea
and
\ba
 &-& (1-\chi)\omega^2 \beta_{C} \tilde \theta_{2}\nonumber \\ &=& {1\over 2} i_{B} \delta (\omega) - {1\over 2} \tilde j + i \omega ( 1 - \rho ) \tilde \theta_{2} - ( 1 - \kappa ) \widetilde{ \sin(\theta_{2})} - i \omega \gamma ( \tilde \theta_{1} - \tilde \theta_{2} ) 
 \ea
Denoting $ \tilde S = \tilde \theta_1 + \tilde \theta_2 $ and $ \tilde D = \tilde \theta_1 - \tilde \theta_2$, we get:
\be
\tilde S = { a (\omega) \kappa + b (\omega) - (i \omega \rho+\omega^2 \chi\beta_C) \tilde D - i_B \delta ( \omega) \over ( \omega^2 \beta_C + i \omega )}
\ee
and
\be
\tilde D = { a ( \omega) + \kappa b (\omega) - (i \omega \rho+\omega^2 \chi\beta_C) \tilde S - \tilde j \over (\omega^2 \beta_C + i \omega + 2 i \omega \gamma)}
\ee
where
$$
a (\omega) = I_1 (\omega) - I_2 (\omega)
$$ and  
$$
b (\omega) = I_1 ( \omega) + I_2 (\omega)
$$
with 
\be
I_1 (\omega) = I \pi ( \delta ( -v_{dc} + \omega ) e^{-i \psi_1} + \delta ( v_{dc} + \omega ) e^ {i \psi_1})\,,
\ee
and 
\be
I_2 (\omega) = I \pi ( \delta ( -v_{dc} + \omega ) e^{-i \psi_2} + \delta ( v_{dc} + \omega ) e^ {i \psi_2})\,.
\ee
Using the notation
\be
s(\omega) = ( \omega^2 \beta_C + i \omega )\,, \quad d(\omega)= A(\omega)-(\omega^2 \beta_C + i \omega + 2 i \omega \gamma)\,,
\ee
\be
r(\omega)=i\rho\omega+\chi\beta_C\omega^2\,,
\ee
\begin{eqnarray}
f_1&=& \kappa a(\omega)+b(\omega)-\delta(\omega) i_b \,,\\ f_2&=&   \beta (a(\omega)+\kappa b(\omega))-8 \pi \phi A(\omega)\delta(\omega)\,,
\end{eqnarray}
we get
\begin{eqnarray}
\tilde S&=&{   \beta d(\omega) f_1 +r(\omega) f_2\over  \beta [d(\omega) s(\omega)-r(\omega)r(-\omega)]}\,,\\
\tilde D &=& { -s(\omega) f_2 + \beta r(\omega) \omega f_1 \over \beta [d(\omega) s(\omega)-r(\omega)r(-\omega)]}\,.
\end{eqnarray}

The inverse Fourier transform leads to
\begin{eqnarray}
S(t)&=& 2 v_{dc}t -{4\rho \phi\over\beta} \nonumber\\&+&{I\over 2}{e^{-i v_{dc} t}\over [d(v_{dc})s(v_{dc})-r(\v)r(-\v)]}\bigg[(1-\kappa)( d(v_{dc})-r(\v))e^{-i\psi_2} \nonumber \\ &&~~~~~~~~~+(1+\kappa)( d(v_{dc})+ r(\v))e^{-i\psi_1}\bigg]+c.c.\,,\\
D(t) &=& {4 \phi \over \beta}-{\rho i_b \beta \over 4} \nonumber\\&+&{I\over 2}{e^{-i v_{dc} t}\over \beta[d(v_{dc})s(v_{dc})-r(\v)r(-\v)]}\bigg[(1-\kappa)(\beta s(v_{dc})+r(\v))e^{i\psi_1}\nonumber \\ &&~~~~~~~~-(1+\kappa)(\beta s(v_{dc})-r(\v))e^{i\psi_2}\bigg]+c.c.\,,\\
\label{j}j(t) &=& {\rm const}\nonumber \\&-&{I\over 2}{e^{-i v_{dc} t} A(v_{dc})\over \beta[d(v_{dc})s(v_{dc})-r(\v)r(-\v)]}\bigg[(1-\kappa)(\beta s(v_{dc})+r(\v))e^{i\psi_1}\nonumber \\ &&~~~~~~~~-(1+\kappa)(\beta s(v_{dc})-r(-\v))e^{i\psi_2}\bigg]+c.c.\,,
\end{eqnarray}
The constant term in $j(t)$ is given by $A(0) (8\phi/\beta -\rho i_b \beta/4)$ but plays no role in the following.
After some tedious algebra it can be shown that (\ref{ibvdc2}) leads to eqn.(\ref{exact}).

\end{document}